\documentclass{pasj01}
\usepackage{bm,url}
\Received{$\langle$reception date$\rangle$}
\Accepted{$\langle$acception date$\rangle$}
\Published{$\langle$publication date$\rangle$}

\begin{document}

\title{ Annual parallax measurement of a Mira variable star BX~Cam with VERA }
\author{
Masako \textsc{Matsuno}$^{1}$, 
Akiharu \textsc{Nakagawa}$^{1}$, 
Atsushi \textsc{morita}$^{1}$, 
Tomoharu \textsc{Kurayama}$^{2}$, 
Toshihiro \textsc{Omodaka}$^{1}$, 
Takumi \textsc{Nagayama}$^{3}$, 
Mareki \textsc{Honma}$^{3, 4}$, 
Katsunori M \textsc{Shibata}$^{3, 4}$, 
Yuji \textsc{Ueno}$^{3}$, 
Takaaki \textsc{Jike}$^{3, 4}$, 
and 
Yoshiaki \textsc{Tamura}$^{3, 4}$
}
\altaffiltext{}{
$^{1}$Graduate School of Science and Engineering, Kagoshima University, 1-21-35 Korimoto, Kagoshima-shi, Kagoshima 890-0065, Japan \\
$^{2}$Teikyo University of Science, 2-2-1 Sakuragi, Senju, Adachi-ku, Tokyo 120-0045, Japan \\
$^{3}$Mizusawa VLBI Observatory, National Astronomical Observatory of Japan, 2-12 Hoshi-ga-oka, Mizusawa-ku, Oshu-shi, Iwate 023-0861, Japan \\
$^{4}$Department of Astronomy, School of Physical Sciences, The Graduate University for Advanced Studies, SOKENDAI, 2-21-1 Osawa, Mitaka, Tokyo 181-8588, Japan \\ 
}
\email{k6908465@kadai.jp}


\maketitle

\begin{abstract}
We report results of astrometric VLBI observations toward a Mira variable star BX~Cam using the VLBI array ``VERA''. 
The observations were performed from February 2012 to November 2014. 
Obtained parallax is 1.73$\pm$0.03 mas corresponding to a distance of 0.58$\pm$0.01 kpc. 
Parallax of this source was also reported in Gaia DR2 as 4.13$\pm$0.25 mas, and there is a 240 \% difference between these two measurements. 
Astrometric results from our VLBI observations show that we exactly traced angular motions of the seven maser spots in BX~Cam. 
We calculated stellar luminosities using both parallaxes, and obtained luminosities of $L_{\ast}^{\mathrm{VERA}} = 4950\pm170 L_{\odot}$ and $L_{\ast}^{\mathrm{Gaia}} = 870\pm110 L_{\odot}$. 
Deduced luminosities also support a validity of the parallax that we determined with VERA. 
Evaluating the two parallaxes, we concluded that the parallax of 1.73$\pm$0.03 mas from the VERA observations is correct for BX~Cam. 
We obtained a systemic motion of BX~Cam as 
(${\mu_{\alpha}}\cos{\delta}^{\mathrm{sys}}$, ${\mu}_{\delta}^{\mathrm{sys}}$) $=$ 
(13.48$\pm$0.14, $-$34.30$\pm$0.18) mas\,yr$^{-1}$.
A total of 73 H$_2$O maser spots detected from our VLBI observations show a spatial distribution of 30 au $\times$ 80 au with a strong elongation along north-south direction. 
They show outflows with a three-dimensional velocity of 14.79$\pm$1.40 km\,s$^{-1}$. 
From a comparison between time variations of $V$-band magnitudes and H$_2$O maser, we found that variation of the H$_2$O maser is relevant to that seen in $V$-band even though the H$_2$O maser does not recover its maximum flux in each cycle. 
\end{abstract}

\section{Introduction}
\label{sec_intro}
Asymptotic giant branch (AGB) stars are cool and bright stars that are at the end of the evolution of low to intermediate mass ($0.8M_{\odot}$ - $8M_{\odot}$) stars. 
Due to their mass loss ratio as high as 10$^{-7}$ to sometimes 10$^{-5} M_{\odot}$\,yr$^{-1}$, the surface material of the stars are released into interstellar space and form planetary nebulae, then finally evolve to white dwarfs with its core.
Mira variable stars are one of the representatives of AGB stars showing typical pulsation periods of 200 to 800 days.
The photosphere is covered with thick dust and molecular gas due to high mass loss phenomena and infrared radiation due to circumstellar envelope is outstanding. 
It is well known that they have SiO, H$_2$O, and OH masers in the dust and molecular gas regions.
Expansion velocity of the outer layers reaches 5 to 10 km\,s$^{-1}$ \citep{kim2014}. 
Among the three maser species, SiO maser is found at close vicinity of the stellar surface. 
The H$_{2}$O maser is found outer region next to the SiO maser, and the OH maser is excited at the outermost region. 
All the maser species are excited in expanding dust shell and trace the stellar outflow. 

BX~Camelopardalis (BX~Cam) is an O-rich Mira variable star with a pulsating period of 486 days (AAVSO)\footnote[1]{The American Association of Variable Star Observers\\https://www.aavso.org}.
It also has names of IRAS 05411$+$6957, IRC$+$70066, NSV 2601, and RAFGL 811. 
The spectral type is M9.5 \citep{solf1987}, and the presence of SiO, H$_{2}$O, and OH masers have been confirmed \citep{1979zuck, cro83, uki84_2}. 
Parallax of BX~Cam is measured by Gaia and 4.13$\pm$0.25 mas is reported in the Gaia Data Release 2 (DR2)\footnote[2]{Gaia Data Release 2; https://www.cosmos.esa.int/web/gaia/dr2}. 
We think it is important to confirm its validity by comparing it with a parallax from an independent measurement based on VLBI astrometry. 
Observation of H$_2$O masers around AGB stars are important not only for parallax measurement but also for understanding of characteristics of circumstellar matters and its kinematics. 

In this paper, we report results of astrometric VLBI observations toward H$_2$O maser in BX~Cam using the VLBI exploration for Radio Astronomy (VERA). 
In section~\ref{sec_Obs}, we describe details of our VLBI observations and single-dish observations at VERA Iriki station. 
In section~\ref{sec_results}, results are presented. 
In section~\ref{sec_dis}, we discuss the validity of parallax measurements. 
Circumstellar kinematics of the maser spots will also be described. 
Finally, our studies are summarized in section~\ref{sec_sum}.

\section{Observation and Data Reduction}
\label{sec_Obs}
\subsection{VLBI observations with VERA}
\label{subsec_vlbiobs}
VLBI observations of BX~Cam were conducted from February 2012 to November 2014 using the VERA.
The VERA array consists of four 20\,m aperture antennas at Mizusawa, Iriki, Ogasawara, and Ishigaki-jima. 
The maximum baseline length of the array is 2270\,km between Mizusawa and Ishigaki-jima stations \citep{kob03}. 
We observed H$_2$O maser emission at the rest frequency of 22.235080\,GHz (6$_{16}$-5$_{23}$ transition) associated with BX~Cam.  
To measure absolute proper motion of H$_2$O maser spots, we determine positions of the maser spots with respect to the position reference source J0554$+$6857. 
Dual-beam system equipped with the VERA antennas enables us to observe these two sources, simultaneously \citep{kaw00}. 
We present coordinates of the two sources on table~\ref{table_coordinate}. 
These coordinates are used as a phase-tracking center in a correlation process of the VLBI data. 
They have separation angle of 1.19$^{\circ}$ with a position angle of 146.8$^{\circ}$. 
In table~\ref{table_obsstatus}, we present observation date and corresponding Modified Julian Date (MJD). 

Received signals at two beams were 2-bit digitized with a sampling ratio of 512 Msps, yielding 1024 Mbps data in each beam. 
The 256 MHz bandwidth data at each beam were divided into 16 IF channels with 16 MHz bandwidth. 
In a recording procedure of maser source BX~Cam, 16 MHz out of 256 MHz data was recorded into a magnetic tape. 
And, in the case of continuum target J0554$+$6857, 240 MHz (16 MHz $\times$ 15 IFs) out of 256 MHz data was recorded into the magnetic tape at the same time. 
In the recording process, the SONY DIR\,2000 system was used. 
Data correlation was made with Mitaka FX correlator \citep{shi98}. 
In the correlation process of BX~Cam, a 16\,MHz data was processed with 512 spectral channels and it gives a frequency spacing of 31.250 kHz with corresponding velocity spacing of 0.42 km\,s$^{-1}$. 
In the correlation process of the reference source J0554$+$6857, each IF channel was processed with 64 spectral channels. 
The synthesized beam size was typically 1.36\,mas$\times$1.00\,mas with its major axis position angle of $-32^{\circ}$ in our observations. 

We used the Astronomical Imaging Package Software (AIPS) developed in the National Radio Astronomical Observatory in the data reduction. 
Amplitude calibration was done using system noise temperatures and gains logged during observations at each station. 
For bandpass calibration of target source BX~Cam, we observed bright continuum sources 0552$+$398 and 3C84 every 80 minutes. 
We used a task FRING to solve residual phase fluctuations of J0554$+$6857 with integration time of 3 minutes. 
By setting software parameters of ``solint = 3'' and ``solsub = 6'' in AIPS, we obtained solutions of phase, group delay, and delay ratio with interval of every 30 seconds. 
And, a task CALIB was also used to solve phase fluctuation with shorter time scales. 
Using a task TACOP, these solutions were transferred to the data of BX~Cam. 
After the calibration, the visibility data were Fourier transformed to phase referenced maps using a task IMAGR. 
Field of view of BX~Cam image was 51.2 mas $\times$ 51.2 mas square (2048 $\times$ 2048 pixel map with its pixel size of 0.025 mas/pixel). 
We made images of several fields, then synthesized them into one wide field image which will be shown in section~\ref{sec_results}. 
Noise levels of the phase referenced map of BX~Cam was, typically, 0.2 Jy\,beam$^{-1}$, and we adopted a signal to noise ratio  (S/N) of 6 as a detection criterion of maser spots. 
On each phase referenced map, we fitted images of the maser spots to two-dimensional Gaussian models to deduce their peak positions and flux densities. 
The positions were used in parallax fitting and used also in analysis of internal motions of the maser spots in BX~Cam. 
More detailed procedures of phase referencing are given in a study by \citet{nak08}. 

\subsection{Single-dish observations at VERA Iriki station}
\label{subsec_singleobs}
In addition to the VLBI observations, we have been conducted long term single-dish monitoring of H$_2$O maser at VERA Iriki station. 
The first single-dish observation of H$_2$O maser in BX~Cam was held in 10 February 2009 at 22 GHz. 
Observation interval has not necessarily been uniform, but in the shortest case, an interval of about one month was adopted. 
This monitoring observation is being continued now. 
The intervals of recent single-dish observations are not so short as the past.  
From an integration time of 10 minutes, 1$\sigma$ noise level of the single-dish observations was obtained to be, typically, 0.05 K corresponding to the flux density of $\sim$1 Jy. 
In the single-dish observations of H$_2$O maser, velocity resolutions of 0.42 km\,s$^{-1}$ and 0.21 km\,s$^{-1}$ were adopted. 
In 2018, we also conducted single-dish SiO maser observation at 43 GHz at VERA Iriki station with a velocity resolution of 0.21 km\,s$^{-1}$. 

\begin{table}[htb]
\caption{Coordinates of the sources}
\begin{center}
\begin{tabular}{lll} \hline 
Source & RA (J2000.0) & DEC (J2000.0)　\\ \hline \hline 
BX~Cam & $05^\mathrm{h}\, 46^\mathrm{m}\, 44.3614^\mathrm{s}$ & $+69^{\circ}\, 58\arcmin\, 24.035\arcsec$ \\
J0554$+$6857 & $05^\mathrm{h}\, 54^\mathrm{m}\, 00.806717^\mathrm{s}$ & $+68^{\circ}\, 57\arcmin\, 54.44379\arcsec$ \\ 
\hline
\end{tabular}
\end{center}
\label{table_coordinate}
\end{table}

\begin{table}
\caption{Date of VLBI observations}
\label{table_obsstatus}
\begin{center}
\begin{tabular}{ccc} 
\hline
Obs. & Date\,\,\,\,\,\,\,\,\,\,\,\, & MJD \\ 
ID & Y\,\,\,\,\,\,\,\,M\,\,\,\,\,D\,& \\ \hline\hline 
1 & 2012~Feb 10 & 55967 \\ 
2 & Mar 09 & 55995 \\
3 & Apr 13 & 56030 \\
4 & Apr 26 & 56043 \\
5 & May 26 & 56073 \\ 
6 & Jun 19 & 56097 \\ 
7 & Aug 22 & 56161 \\ \hline
8 & 2013~Jan 17 & 56309 \\ 
9 & Feb 25 & 56348 \\
10 & Mar 24 & 56375 \\
11 & Apr 30 & 56412 \\
12 & May 26 & 56438 \\
13 & Sep 01 & 56536 \\
14 & Oct 21 & 56586 \\
15 & Nov 22 & 56618 \\
16 & Dec 20 & 56646 \\ \hline
17 & 2014~Jan 24 & 56681 \\
18 & Mar 13 & 56729 \\
19 & Apr 23 & 56770 \\
20 & May 26 & 56803 \\
21 & Sep 05 & 56905 \\
22 & Oct 26 & 56956 \\
23 & Nov 24 & 56985 \\
\hline
\end{tabular}
\end{center}
\end{table}

\section{Results}
\label{sec_results}
\subsection{Spectra of H$_2$O and SiO masers obtained from single-dish observations}
\label{subsec_spectr}
\begin{figure}[htbp]
\begin{center}
\includegraphics[width=80mm, angle=0]{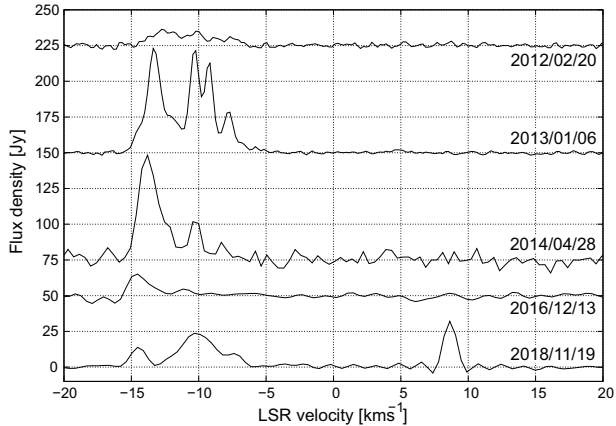} 
\end{center}
\caption{
Total power spectra of H$_{2}$O maser in BX~Cam obtained at VERA Iriki station from 20 February 2012 to 19 November 2018. 
For clarity, noise floors are shifted appropriately with several $\times$10 Jy.
}
\label{fig_single}
\end{figure}

We obtained H$_{2}$O maser spectra and its time variation in BX~Cam from the single-dish observations at Iriki station. 
Figure~\ref{fig_single} shows five total power spectra observed from 20 February 2012 to 19 November 2018.
Noise floors of each spectrum is shifted to {several $\times$ 10 Jy for clarity}. 
During the period of our VLBI observations, from February 2012 to November 2014, the H$_{2}$O maser have continuously been found at $V_{\mathrm{LSR}}$ range in between $-$7 km\,s$^{-1}$ to $-$15 km\,s$^{-1}$. 
We detected multiple velocity components in all single-dish observations. 
In 2018, we detected red-shifted component at the LSR velocity ($V_{\mathrm{LSR}}$) of $+$7 km\,s$^{-1}$ to $+$10 km\,s$^{-1}$ in single-dish observation, however it was not detected during our VLBI observations from 2012 to 2014. 
We also confirmed existence of the same red-shifted components in a study by \cite{En1988}. 
After the last detection in March 2019 at Iriki station, the red-shifted components have been disappeared again. 

\begin{figure}[htbp]
\begin{center}
\includegraphics[width=80mm, angle=0]{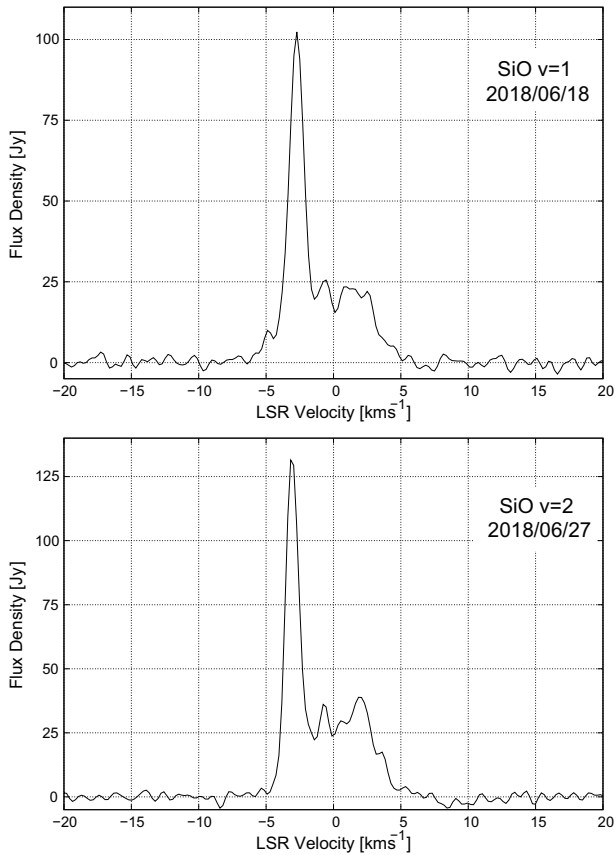} 
\end{center}
\caption{
Total power spectra of SiO maser v=1 (top) and v=2 (bottom) in BX~Cam obtained at VERA Iriki station in 18 and 27 June, 2018, respectively.
SiO maser was detected in radial velocity range of $-$5 km\,s$^{-1}$ to $+$5 km\,s$^{-1}$.
}
\label{fig_SiO_v1v2}
\end{figure}

Figure~\ref{fig_SiO_v1v2} shows total power spectra of SiO maser observed at Iriki station. 
Top and bottom panels are SiO v=1 and SiO v=2 obtained in 18 and 27 June 2018, respectively. 
Bright emissions with a velocity range from $-5$ km\,s$^{-1}$ to $+5$ km\,s$^{-1}$ were detected. 
Since the excitation of SiO maser occur at the close vicinity of the central star rather than H$_2$O maser, velocity of the SiO maser emission is thought to reflect radial velocity of the central star. 
From our results, we can estimate the radial velocity of BX~Cam to be 0 km\,s$^{-1}$, and this is consistent with the radial velocity of 0 km\,s$^{-1}$ reported in studies by \citet{Nyman1992} and \citet{kim2010}. 

\subsection{Annual parallax and Distance of BX~Cam} 
\label{sec_parallax}
Among a total of 23 VLBI observations, we detected H$_{2}$O maser in 21 observations using a technique of phase referencing analysis. 
We adopted a signal to noise ratio (S/N) of 6 as a detection criterion of maser spots in phase referenced images. 
The bluest and reddest LSR velocities ($V_{\mathrm{LSR}}$) of the detected maser spots were $-$15.69 km\,s$^{-1}$ and $-$7.29 km\,s$^{-1}$, representing a velocity coverage of 8.40 km\,s$^{-1}$. 
We detected maser spots in all 21 discrete velocity channels in the velocity coverage. 

In data reductions of the two observations on 1st September 2013 and 5 September 2014, we did not obtain valid solutions for calibration. 
Therefore, we could not make phase referenced images of maser spots. 
In the second observation in 09 March 2012, we could not identify the maser spots that we trace for parallax estimation. 
There is a significantly large discontinuity in position of the maser spots. 
Finally, we used 21 observations to solve an annual parallax of BX~Cam. 
Table~\ref{table_maser} represents detection flags of the maser spots in order of increasing LSR velocities ($V_{\mathrm{LSR}}$).
The flags of ``1'' and ``0'' correspond to detection and non-detection of the maser spots, respectively. 

\begin{figure}[htbp]
\begin{center}
\includegraphics[width=80mm, angle=0]{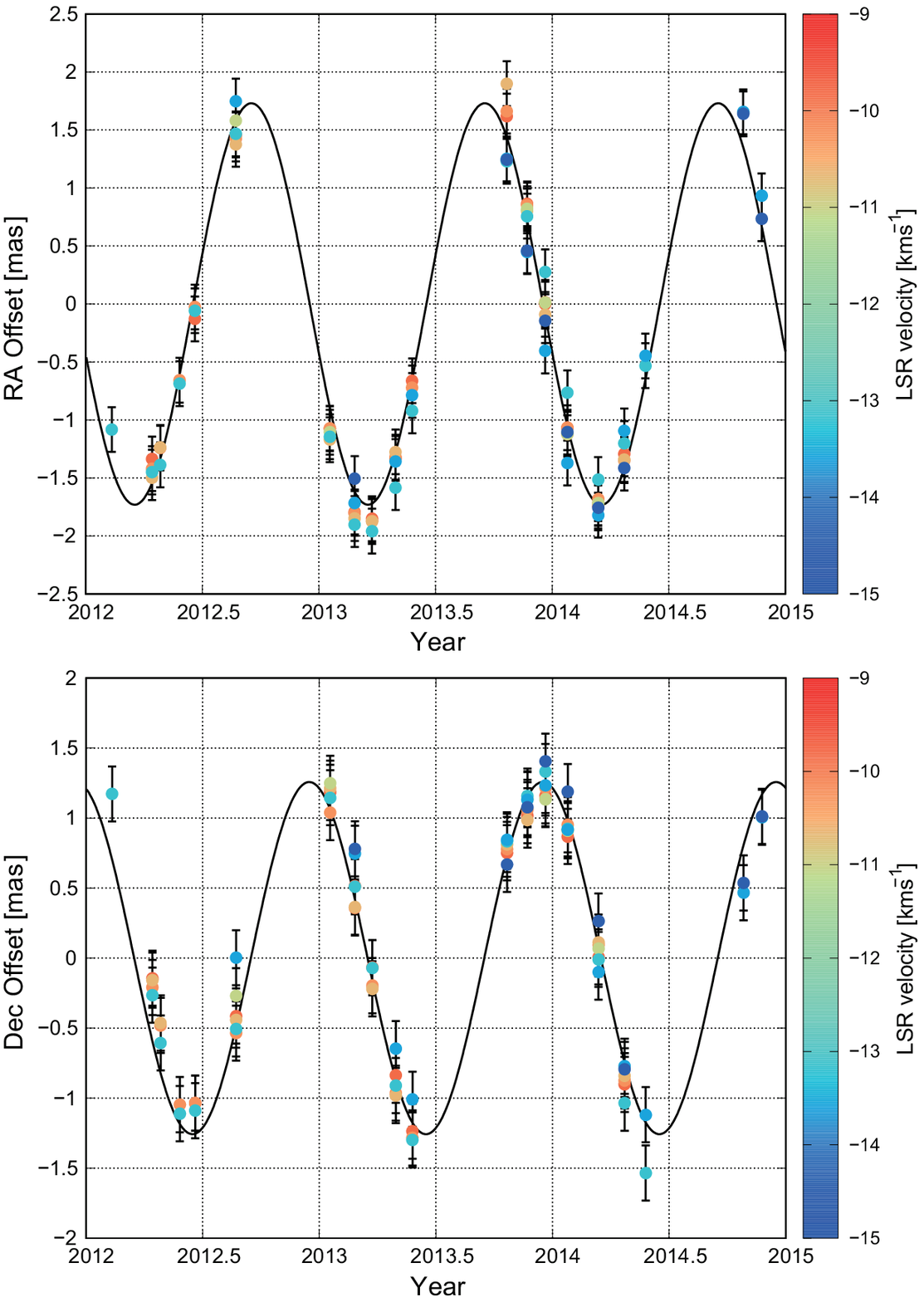} 
\end{center}
\caption{
Oscillating terms of the maser spots in BX~Cam along the axes of RA (top) and DEC (bottom). 
Filled circles and solid lines represent positions of the maser spots and best-fit models of the annual parallax. 
Color indicate LSR velocities ($V_{\mathrm{LSR}}$) of each maser spot. 
}
\label{fig_parallax}
\end{figure}

To solve an annual parallax, it is important to select maser spots with simple and stable structures. 
Using seven maser spots which represent point like structure, we estimated the parallax of BX~Cam. 
In table~\ref{table_maser}, we appended daggers ($\dag$) to the $V_{\mathrm{LSR}}$ of maser spots which are used in the parallax fitting. 
The observation IDs are same as those presented in table~\ref{table_obsstatus}. 
Using a least-squares analysis, a parallax of BX~Cam was obtained to be 1.73$\pm$0.03 mas which corresponds to the distance of 0.58$\pm$0.01 kpc.
In figure~\ref{fig_parallax}, we presented oscillating terms of parallactic motion in RA (top panel) and DEC (bottom panel) along with time. 
Filled circles are position offsets of each maser spot and their colors indicate the $V_{\mathrm{LSR}}$. 
Solid curves are the best-fit models of the parallactic oscillations.  
We regarded post-fit residuals of the least-squares analysis, 0.19 mas in RA axis and 0.18 mas in DEC axis, as position errors of the maser spots and presented them as error bars in figure~\ref{fig_parallax}. 
For three maser spots among all the seven spots which were used in the parallax fitting, selection of the maser spots representing point-like structure, and identification of the same spots were relatively difficult. 
And also, residual four maser spots were found in very adjacent positions aligned in the direction of east-west with angular separations smaller than the size of synthesized beam. 
We think that these conditions can introduce relatively large post-fit residual in RA axis. 
In the parallax fitting, proper motions of each maser spot are simultaneously solved. 
We present them in the next section. 

\begin{table*}[]
\caption{Maser spots of BX~Cam detected in phase referencing analysis.} 
\begin{center}
\label{table_maser}
\begin{tabular}{lccccccccccccccccccccc} 
\hline
& \multicolumn{21}{c}{Observation ID} \\ \cline{2-22} $V_{\mathrm{LSR}}$ & & & & & & & & & & & & & & & & & \\ 

[km\,s$^{-1}$] &1&2&3&4&5&6&7&8&9 &10&11&12&14&15&16&17&18&19&20&22&23 \\ \hline\hline
$-$7.29&0&0&0&1&1&0&1&1&1&0&0&0&0&0&0&0&0&0&0&0&0\\ 
$-$7.71&1&0&0&1&1&1&1&1&1&1&0&0&0&0&1&0&0&0&0&0&0\\ 
$-$8.13&0&0&0&0&0&0&1&1&1&0&0&0&0&0&0&0&0&0&0&0&0\\
$-$8.55&0&0&0&0&0&0&1&1&0&0&0&0&0&0&0&0&0&0&0&0&0\\ 
$-$8.97&0&0&0&0&0&0&1&1&1&1&0&0&1&1&0&1&0&0&0&0&0\\
$-$9.39&0&0&0&0&0&0&1&1&1&1&0&0&1&1&1&1&0&0&0&0&0\\ 
$-$9.81$^{\dag}$ &1&1&1&1&0&1&1&1&1&1&1&1&1&1&1&1&1&1&0&0&0\\ 
$-$10.23$^{\dag}$ &1&1&1&1&1&1&1&1&1&1&1&1&1&1&1&1&1&1&0&0&0\\ 
$-$10.65$^{\dag}$ &0&0&1&1&0&1&1&1&1&1&1&0&1&1&1&1&1&1&0&0&0\\ 
$-$11.07$^{\dag}$ &1&0&0&0&0&0&1&1&0&0&0&0&0&1&1&1&1&0&0&0&0\\ 
$-$11.49&1&0&0&1&0&0&0&1&1&1&0&0&0&1&1&1&1&1&0&0&0\\ 
$-$11.91&1&0&1&1&1&0&0&1&1&0&0&0&0&1&1&1&1&1&1&0&0\\ 
$-$12.33&1&0&1&1&1&1&1&1&0&0&0&0&0&1&1&1&1&1&1&0&0\\ 
$-$12.75&1&1&1&1&1&1&1&1&1&1&1&0&1&1&1&1&0&1&0&0&0\\ 
$-$13.17$^{\dag}$ &1&1&1&1&1&1&1&1&1&1&1&1&1&1&1&1&1&1&1&0&0\\ 
$-$13.59$^{\dag}$ &0&1&0&1&1&0&1&0&1&1&1&1&1&1&1&1&1&1&1&1&1\\ 
$-$14.01&0&0&0&0&0&0&0&1&1&0&0&0&1&1&1&1&1&1&1&1&1\\ 
$-$14.43$^{\dag}$ &0&0&0&0&0&0&0&1&1&1&0&0&1&1&1&1&1&1&0&1&1\\ 
$-$14.85&0&0&0&0&0&0&0&1&0&0&0&0&1&1&1&1&1&1&1&1&1\\ 
$-$15.27&0&0&0&0&0&0&0&0&0&0&0&0&0&1&1&1&1&1&0&0&0\\ 
$-$15.69&0&0&0&0&0&0&0&0&0&0&0&0&0&1&0&1&1&1&0&0&0\\ 
\hline
\multicolumn{1}{@{}l@{}}{\hbox to 0pt{\parbox{140mm}{\footnotesize
\smallskip
${\dag}$: The $V_{\mathrm{LSR}}$ appended with daggers indicate that the velocity components were used in parallax fitting. 
}\hss}} 
\end{tabular}
\end{center}
\end{table*}

\subsection{Distribution and proper motion of maser spots} 
In the phase referenced images, we detected a total of 73 maser spots within a velocity coverage from $-$15.69 km\,s$^{-1}$ to $-$7.29 km\,s$^{-1}$. 
Since we found multiple maser spots at same velocity channels, total number of the maser spots are larger than the number of velocity channels. 
We describe details of the 23 maser spots in table~\ref{table_maserdistribution} from which we could derive proper motions. 
In the case where a maser spot is detected in less than two continuous observations, we do not define proper motion, and subsequently there is no description of the maser spot at the $V_{\mathrm{LSR}}$ of $-$8.55 km\,s$^{-1}$ in table~\ref{table_maserdistribution}. 
Since we found two discrete maser spots in the three velocity channels of $-8.97$ km\,s$^{-1}$, $-11.49$ km\,s$^{-1}$, and $-12.75$ km\,s$^{-1}$, there are two positions and proper motions for each maser spot in table~\ref{table_maserdistribution}. 
In the table, ${\Delta\alpha}\,\cos {\delta}$ and ${\Delta\delta}$ represent position of the maser spot with respect to the phase tracking center of BX~Cam. 
The ${\mu_{\alpha}}\cos{\delta}$ and ${\mu}_{\delta}$ are angular proper motion of the maser spot in units of mas\,yr$^{-1}$. 
The $S$ and S/N are flux density and signal to noise ratio, respectively. 
The positions and flux densities of each maser spot in table~\ref{table_maserdistribution} are measurements at the time of their first detections. 
For the seven maser spots used in the parallax fitting, one common parallax and proper motions of each maser spot are solved at the same time. 
For the other maser spots, proper motions are solved from least-square analysis while using a fixed parallax of 1.73 mas. 

\begin{figure}[htbp]
\begin{center}
\includegraphics[width=80mm, angle=0]{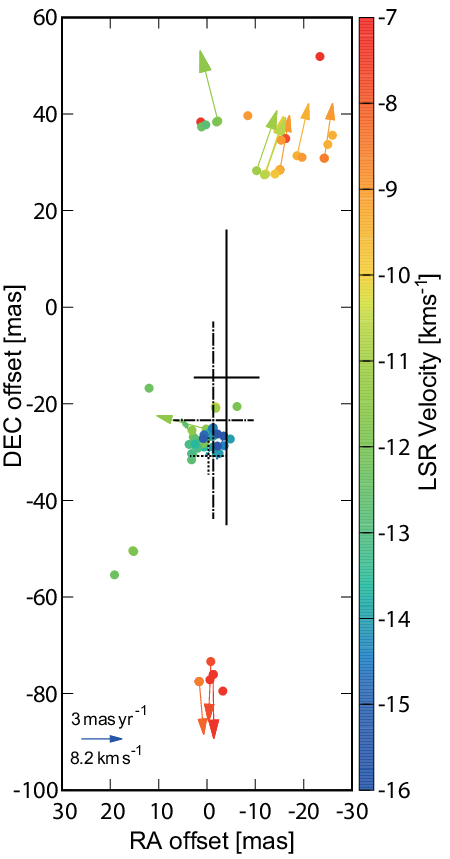} 
\end{center}
\caption{
Distribution and internal motions of maser spots in BX~Cam. 
Filled circles and arrows represent maser spots and their internal motions. 
LSR velocities ($V_{\mathrm{LSR}}$) are indicated with colors. 
The stellar velocity of BX~Cam is 0 km\,s$^{-1}$. 
The arrow at the bottom left corner represents a motion of 3 mas\,yr$^{-1}$ which corresponds to a linear velocity of 8.2 km\,s$^{-1}$ at the distance of 0.58 kpc. 
Crosses with solid line, one-dotted chain line, and dotted-line represent stellar positions estimated from data sets of VERA and Gaia. 
More details are seen in section~\ref{subsec_masermotion}. 
}
\label{fig_maserdistribution}
\end{figure}

Figure~\ref{fig_maserdistribution} shows distribution of all 73 maser spots on the sky plane. 
The distribution is highly elongated along north-south direction with an angular extent of 50 mas $\times$ 140 mas which corresponds to 30 au $\times$ 80 au at the distance of BX~Cam. 
It can suggest a presence of collimated flows. 
Filled circles indicate maser spots and color indicate their $V_{\mathrm{LSR}}$. 
From the distribution, we find that the red-shifted components mainly distribute in northern and southern area of the whole extent, and the blue-shifted components are well clustered and distribute in the central region. 
Details of the arrows which indicate circumstellar motion of the maser spots are given in section~\ref{subsec_masermotion}. 
At bottom left of figure~\ref{fig_maserdistribution}, we presented a vector of 3 mas\,yr$^{-1}$ which corresponds to a linear velocity of 8.2 km\,s$^{-1}$ at 0.58 kpc, 

\begin{table*}[]
\caption{Positions and proper motions of the maser spots in BX~Cam. }
\begin{center}
\begin{tabular}{ccccccccl} 
\hline
Spot & $V_{\mathrm{LSR}}$ & ${\Delta\alpha}\,\cos {\delta}$& ${\Delta\delta}$& ${\mu_{\alpha}}\cos{\delta}$& ${\mu}_{\delta}$& $S$& S/N\\ 
ID & [km s$^{-1}$]&[mas] & [mas] & [mas\,yr$^{-1}$] & [mas\,yr$^{-1}$] & [Jy\,beam$^{-1}$] \\ \hline\hline
1 & $-$7.29 & $-$1.24 & $-$76.03 & 13.42$\pm$0.24 & $-$39.05$\pm$0.29 & 0.90 & 8.1 \\
2 & $-$7.71 & $-$0.80 & $-$73.36 & 13.65$\pm$0.07 & $-$38.75$\pm$0.14 & 1.84 & 14.3 \\
3 & $-$8.13 & 1.62 & $-$77.50 & 13.14$\pm$0.13 & $-$38.21$\pm$0.21 & 2.77 & 12.3 \\
4 & $-$8.97 & $-$15.10 & 28.40 & 12.78$\pm$0.08 & $-$30.26$\pm$0.14 & 12.6 & 16.8 \\
5 & $-$8.97 & $-$24.29 & 30.80 & 12.88$\pm$0.09 & $-$30.22$\pm$0.11 & 6.7 & 11.6 \\
6 & $-$9.39 & $-$18.62 & 31.34 & 12.62$\pm$0.13 & $-$30.45$\pm$0.12 & 18.3 & 42.9 \\
7 & $-$9.81 & $-$12.09 & 27.56 & 12.15$\pm$0.08 & $-$30.03$\pm$0.09 & 1.11 & 7.2 \\
8 & $-$10.23 & $-$11.93 & 27.52 & 12.00$\pm$0.07 & $-$30.12$\pm$0.07 & 2.17 & 12.8 \\
9 & $-$10.65 & $-$11.84 & 27.35 & 11.90$\pm$0.08 & $-$30.06$\pm$0.08 & 1.03 & 7.8 \\
10 & $-$11.07 & $-$10.25 & 28.29 & 11.98$\pm$0.14 & $-$29.88$\pm$0.14 & 3.92 & 12.2 \\
11 & $-$11.49 & $-$2.27 & 38.53 & 14.81$\pm$0.50 & $-$29.10$\pm$0.15 & 3.34 & 13.7 \\
12 & $-$11.49 & 0.17 & $-$25.23 & 17.13$\pm$0.31 & $-$33.31$\pm$0.29 & 0.99 & 7.3 \\
13 & $-$11.91 & 2.83 & $-$26.88 & 14.37$\pm$0.05 & $-$33.00$\pm$0.16 & 1.13 & 11.0 \\
14 & $-$12.33 & 2.67 & $-$27.12 & 14.36$\pm$0.03 & $-$32.92$\pm$0.13 & 1.89 & 18.0 \\
15 & $-$12.75 & $-$1.21 & $-$26.67 & 14.61$\pm$0.11 & $-$34.64$\pm$0.09 & 2.19 & 14.4 \\
16 & $-$12.75 & 2.37 & $-$27.39 & 14.36$\pm$0.09 & $-$32.98$\pm$0.23 & 1.82 & 10.3 \\
17 & $-$13.17 & $-$1.24 & $-$26.74 & 14.38$\pm$0.06 & $-$34.53$\pm$0.06 & 1.59 & 12.3 \\
18 & $-$13.59 & 0.62 & $-$28.88 & 14.39$\pm$0.09 & $-$34.39$\pm$0.09 & 5.85 & 10.3 \\
19 & $-$14.01 & $-$1.31 & $-$28.93 & 14.08$\pm$0.22 & $-$35.00$\pm$0.14 & 17.1 & 16.8 \\
20 & $-$14.43 & $-$3.47 & $-$28.68 & 14.10$\pm$0.13 & $-$34.55$\pm$0.13 & 1.99 & 10.2 \\
21 & $-$14.85 & $-$1.29 & $-$25.03 & 13.75$\pm$0.10 & $-$34.99$\pm$0.19 & 1.29 & 13.3 \\
22 & $-$15.27 & $-$1.12 & $-$26.44 & 13.28$\pm$0.20 & $-$34.17$\pm$0.11 & 4.14 & 28.8 \\
23 & $-$15.69 & $-$3.46 & $-$26.69 & 13.99$\pm$0.12 & $-$36.00$\pm$0.78 & 1.99 & 17.9 \\
\hline
\end{tabular}
\label{table_maserdistribution}
\end{center}
\end{table*}

\section{Discussion}
\label{sec_dis}
\subsection{Time variation of H$_2$O maser in BX~Cam}
\label{subsec_timescale}
\begin{figure}[htb]
\begin{center}
\includegraphics[width=80mm, angle=0]{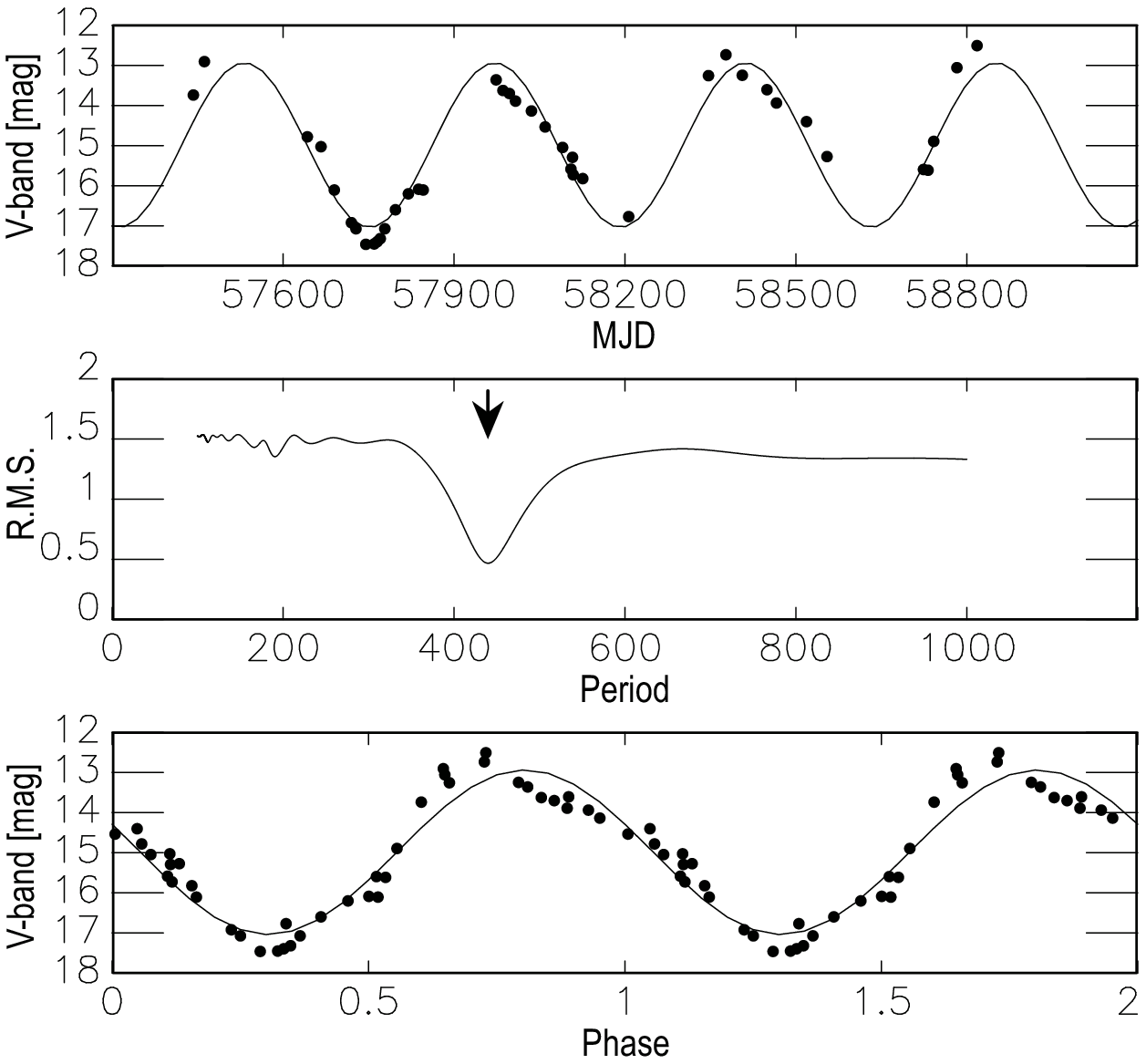} 
\end{center}
\caption{
Period determination using $V$-band light curve. 
Top panel shows a variation of apparent $V$-band magnitude (filled circles) along the MJD. 
Solid curve is the best fit model obtained in the fitting. 
Middle panel shows r.m.s. values along trial periods in Fourier fitting program. 
The period with the smallest r.m.s. is represented by an arrow. 
Bottom panel is the plot of phase versus the apparent magnitude. 
Filled circles indicate $V$-band magnitude same as the top panel. 
}
\label{fig_fourier}
\end{figure}

It is known that an excitation of H$_2$O maser is sensitive to stellar wind due to a stellar pulsation and there is a correlation between time variation of H$_2$O maser and optical variability with 0.1 to 0.4 phase shift relative to optical maximum \citep{kim2014}. 
The optical period of BX~Cam is reported to be 486 days (AAVSO), however we derived the period again by ourselves. 
For BX~Cam, the $V$-band observations were performed on 40 nights in AAVSO database. 
First and last dates of the nights are 25 February 2016 (MJD 57443) and 1st December 2019 (MJD 58818), respectively. 
Compiling the data set of $V$-band magnitudes in 40 nights from AAVSO database, we obtained a pulsation period and a mean magnitude as 439.83 days and 15.00$\pm$0.47 mag, respectively. 
The fitting program was provided from N. Matsunaga (2008, private communication) in which the best-fit Fourier component was determined from many trials. 
The result is shown in figure~\ref{fig_fourier}. 
A top panel in figure~\ref{fig_fourier} shows a variation of $V$-band magnitude (filled circles) with the best fit model (solid curve). 
Middle panel shows root mean square (r.m.s.) of the residuals along with trial periods. 
At the best-fit pulsation period, the r.m.s. shows minimum value. 
In the bottom panel, folded representation of $V$-band magnitudes (filled circles) on pulsation phase is shown. 

\begin{figure}[htb]
\begin{center}
\includegraphics[width=80mm, angle=0]{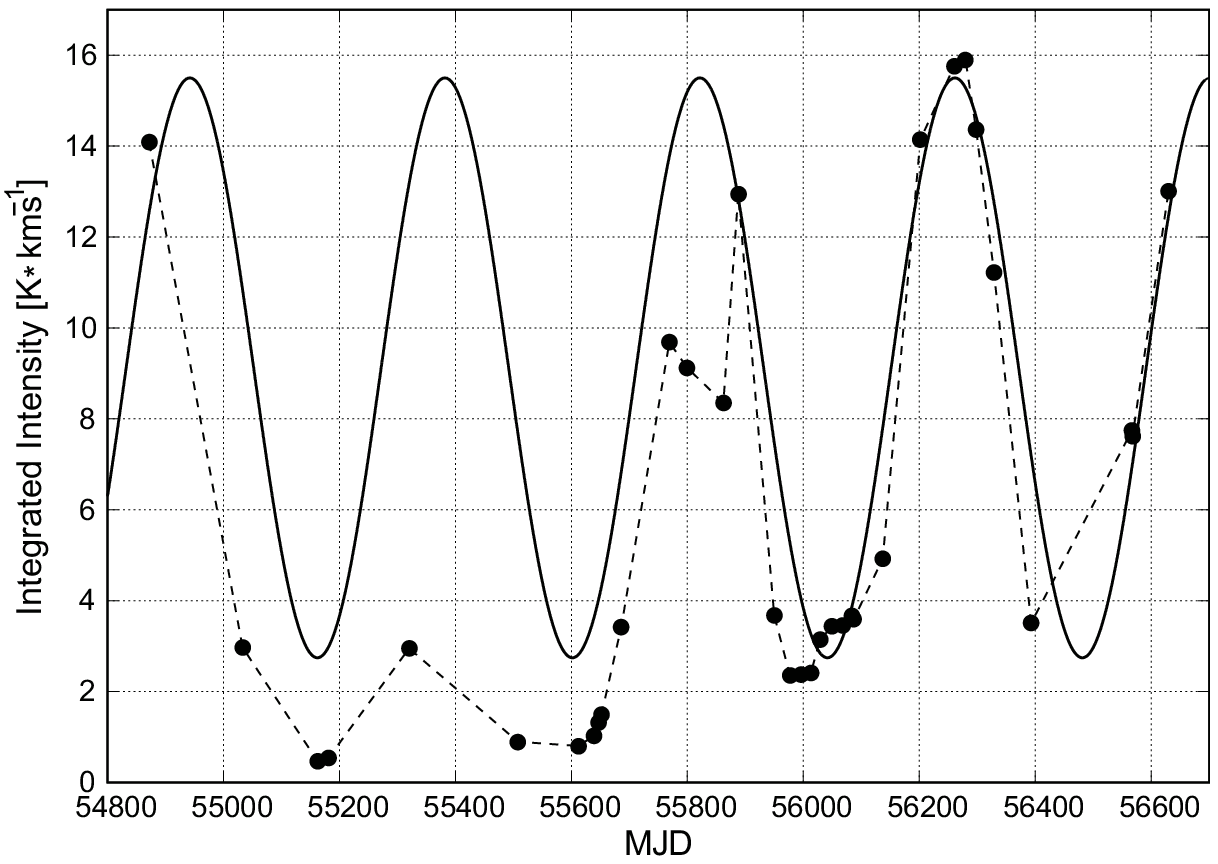} 
\end{center}
\caption{
Integrated intensity of H$_2$O maser (filled circles) obtained from single-dish observations at Iriki station. 
Solid line is the best fit model of the pulsation period obtained from our own fitting using $V$-band monitoring data in AAVSO. 
}
\label{fig_maserlightcurve}
\end{figure}

Now we consider a time variation of H$_2$O maser by using the pulsation period of $P =$ 439.83 days which was derived from our previous fitting. 
To grasp a total activity of the maser emissions in BX~Cam, we use an integrated intensity ($I$) in unit of K$\ast$km\,s$^{-1}$ as an integration of flux density $S$ along the axis of radial velocity. 
In the fitting of phase of H$_2$O maser, we assumed a simple sinusoidal function as 
\begin{displaymath}
I = \Delta I \sin (2 \pi \frac{T}{P} + \theta ) + I_0, 
\end{displaymath}
where $\Delta I$, $T$, $P$, $\theta$, and $I_0$ are amplitude, time, period, phase, and mean intensity, respectively. 
We solved only one phase parameter $\theta$ from the fitting. 
The three parameters of $\Delta I$, $I_0$, and $P$ were fixed. 
Because of large time variation of the integrated intensity, we assumed $\Delta I = 6.38$ K$\ast$km\,s$^{-1}$ and $I_0 = 9.12$ K$\ast$km\,s$^{-1}$ as an amplitude and mean magnitude of H$_2$O maser. 
Both values were estimated from integrated intensities obtained in the time range from MJD 55800 to MJD 56600 where number of single-dish observations is relatively dense.  
Since there are no $V$-band magnitudes in the AAVSO database during the time range of our single-dish observation, we could not directly compare the variations in $V$-band and H$_2$O maser.
In figure~\ref{fig_maserlightcurve}, we present results of the fitting. 
Filled circles indicate integrated intensities of H$_2$O maser observed at VERA Iriki station from 10 February 2009 (MJD 54872) to 4th December 2013 (MJD 56630). 
Solid line indicates the best-fit model. 
It is shown that, over two cycles of pulsation, the local maximum and minimum of the H$_2$O maser synchronize with the variation of $V$-band magnitude even though the H$_2$O maser does not always recover its whole amplitude in each cycle. 
From this consideration, we can expect that pulsation periods of AGB stars can also be derived from the time variation of H$_2$O maser for sources those who have low optical magnitudes but have bright maser emission. 
For example, in observations of OH/IR stars who show low magnitudes in the optical band and sometimes very weak even in the infrared band, time variation of the maser emission can help to determine pulsation periods of the stars. 

\subsection{Evaluation of the parallaxes obtained from VERA and Gaia based on stellar luminosity}
\label{sec_L}
We derived an annual parallax of 1.73$\pm$0.03 mas from astrometric VLBI observations using the VERA, while we can also find a parallax of the same source as 4.13$\pm$0.25 mas in the Gaia Data Release 2 (DR2). 
There is a large discrepancy between the two measurements, representing a difference of 240 \%, however both measurements give relative errors less than 10 \%. 
VERA and Gaia DR2 give relative errors of 2 \% and 6 \%, respectively. 
In this section, we focus on the difference of two measurements and evaluate the parallaxes based on stellar luminosities deduced from the different parallaxes. 

It is difficult to give a clear explanation of the reason why the two measurements show such a large discrepancy, nevertheless we infer the reason by considering characteristics of the target source and observation techniques. 
In VLBI observations, we do not detect any emission from the central star but make images of maser spots associated with the central star. 
By tracing transition of the maser positions, we solve the parallax. 
The angular resolution of the VLBI method, typically it is $\sim$1 mas, is comparable with the angular size of the maser spots. 
If the spatial structure of the maser spot is simple and stable, and also the phase referencing calibration is ideally performed, the deduced parallax is thought to be affected only by thermal noise of the final image. 
As we can see in figure~\ref{fig_parallax}, phase referencing observations were successfully done, and angular motions of the seven maser spots were exactly traced by the astrometric observations with VERA. 

In the Gaia project, parallax is determined from optical observations, and its measurement is performed by tracing transition of optical centroid of the star. 
For AGB stars, size of the stellar photosphere is typically several au and consequently angular size of the photosphere is same as the size of the parallax. 
It is understood that surface brightness distribution and color of AGB star is variable, so it results in a shift of the optical centroid that the Gaia measures. 
This can add significant uncertainty to the positions from which the Gaia parallax is solved \citep{lan18, Chia2018}. 
And it can also give incorrect parallax even though the calibration is ideally performed. 

Since BX~Cam is an AGB star, it is not surprising that VLBI and Gaia give different parallaxes, however this time both parallaxes are determined to a high degrees of accuracies. 
Next, we will consider stellar luminosities to evaluate the parallax of BX~Cam.
In order to estimate a stellar luminosity of BX~Cam ($L_{\ast}$), we use a bolometric correction reported by \citet{bes84}. 
They give a relation between the bolometric correction $BC_K$ and a color $(V-K)$ as, $BC_K = [20.66 - 19.2 (V-K)] / [1.0 - 5.5 (V-K)]$. 
To obtain a color of BX~Cam, we need apparent magnitudes in $V$-band ($m_V$) and $K$-band ($m_K$). 
We adopted $m_V =$ 15.00 (AAVSO) and $m_K = $ 1.05 \citep{oli01} and obtained a color $(V-K) = (m_V - m_K) = 13.95.$ 
Then the $BC_K$ was obtained to be 3.26. 
Using these values, an apparent bolometric magnitude of BX~Cam ($m_{\mathrm{bol}}$) was found to be $m_{\mathrm{bol}} = m_K + BC_K = 4.31$. 
This $m_{\mathrm{bol}}$ can be converted to an absolute bolometric magnitude $M_{\mathrm{bol}}$ of $-4.50$ by applying the parallax of 1.73 mas derived from VERA observation. 
We finally obtained $L_{\ast}^{\mathrm{VERA}} = 4950\pm170 L_{\odot}$ using a relation between luminosity and bolometric magnitude, $M_{\mathrm{bol}} = 4.74 - 2.5 \log (L_{\ast} / L _{\odot})$, where 4.74 is an absolute bolometric luminosity of the sun. 
Following the same procedure, we calculated the luminosity of BX~Cam by applying another parallax of 4.13 mas derived from Gaia observation. 
Then we obtained a luminosity $L_{\ast}^{\mathrm{Gaia}} = 870\pm110 L_{\odot}$. 
In both luminosity estimations, errors are deduced by considering parallax errors in VERA and Gaia measurements. 
Generally, red giants with a mass of 1$M_{\odot}$ have a luminosity of $\sim10^{4} L _{\odot}$, and red supergiants with a mass of 10 $M_{\odot}$ have a luminosity of $\sim10^5 L _{\odot}$ \citep{Eli1992}. 
And also, \citep{tak13} reported luminoisties of a dozen of Mira variables stars representing luminosities of $\sim10^3 L _{\odot}$. 
Comparing the two estimations of stellar luminosities, we can see that $L_{\ast}^{\mathrm{VERA}} = 4950\pm170 L_{\odot}$ is more applicable as a luminosity of Mira variable star. 

Accordingly, we conclude here that the VERA gave correct parallax measurement rather than the Gaia DR2. 
This study shows that VLBI astrometry still provides very valuable data in the distance determination and quantitative studies of Mira variables and other AGB stars. 

\subsection{Dynamics of circumstellar matter}
\label{subsec_masermotion}
As already shown in figure~\ref{fig_maserdistribution}, we presented distribution of the maser spots in BX~Cam. 
In this section, we will reveal dynamics of the circumstellar masers based on our long term VLBI observations. 
The proper motions presented in table~\ref{table_maserdistribution} are interpreted as a composed vector of a systemic motion of the stellar system of BX~Cam and internal motions of each maser spot with respect to the central star. 
Therefore, it is necessary to estimate the systemic motion of BX~Cam to derive internal motions of the maser spots. 

To estimate the systemic motion of BX~Cam, we adopted a procedure like that used in the study by \citet{nak14}. 
They consider “maser group” to derive a systemic motion of a stellar system. 
Distribution of H$_{2}$O maser spots in BX~Cam was not isotropic but shows a bipolar distribution splitting into three major parts. 
We defined three groups as a cluster of maser spots and denoted them as North, Center, and South groups. 
Constituents of each maser group represent narrow coverage of $V_{\mathrm{LSR}}$ and fall adjacent to each other on the sky plane. 
We took averages of proper motions of the spots belonging to each group and presented the three group motions in table~\ref{table_group}. 
Averaging the three group motions again, we finally obtained a systemic motion of BX~Cam as 
(${\mu_{\alpha}}\cos{\delta}^{\mathrm{sys}}$, ${\mu}_{\delta}^{\mathrm{sys}}$)$_{\mathrm{VERA}}$ $=$ 
(13.48$\pm$0.14, $-$34.30$\pm$0.18) mas\,yr$^{-1}$. 
We subtracted this systemic motion from proper motions of 23 maser spots in table~\ref{table_maserdistribution} then obtained internal motions which are presented as vectors in figure~\ref{fig_maserdistribution}. 
We can see that the distribution and internal motion of maser spots indicate bipolar outflow. 

\begin{table}[htb]
\caption{Proper motion of maser groups}
\begin{center}
\begin{tabular}{lcc} \hline 
& $\mu_{\mathrm{x}}$ & $\mu_{\mathrm{y}}$ \\ 
Group & [mas\,yr$^{-1}$] & [mas\,yr$^{-1}$] \\ \hline \hline
North & 12.64$\pm$0.15 & $-$30.02$\pm$0.11\\
Center& 14.40$\pm$0.13 & $-$34.21$\pm$0.20\\
South& 13.40$\pm$0.15 & $-$38.67$\pm$0.21\\ \hline
Average & 13.48$\pm$0.14 & $-$34.30$\pm$0.18 \\ \hline 
\end{tabular}
\end{center}
\label{table_group}
\end{table}

In the Gaia DR2, we can find a systemic motion of BX~Cam as 
(${\mu_{\alpha}}\cos{\delta}^{\mathrm{sys}}$, ${\mu}_{\delta}^{\mathrm{sys}}$)$_{\mathrm{Gaia}}$ $=$ 
(15.53$\pm$0.31, $-$33.66$\pm$0.32) mas\,yr$^{-1}$. 
Residual of the two measurements of proper motion is 2.42 mas\,yr$^{-1}$ and $-$0.24 mas\,yr$^{-1}$ in RA and DEC, respectively. 
This residual correspond to linear velocities of 6.6 km\,s$^{-1}$ and 0.7 km\,s$^{-1}$ at the source distance of 0.58 kpc. 
Generally, the H$_2$O maser around Mira variable stars move outward at an expansion velocity of $\sim$10 km\,s$^{-1}$ \citep{BJ1994}. 
In a study by \citet{kim2010} about velocity profiles of SiO and H$_2$O masers in BX~Cam, the radial velocity of $-$0.9 km\,s$^{-1}$ for both SiO v$=$1 and v$=$2 are reported. 
The radial velocity of $-$10.7 km\,s$^{-1}$ for H$_2$O maser is also reported. 
In a study of ciucumstellar CO emission by \citet{Nyman1992}, a radial velocity of 0 km\,s$^{-1}$ and an expansion velocity of 22.0 km\,s$^{-1}$ are reported for BX~Cam. 
Since in the VLBI observations, positions and motions of circumstellar maser spots are measured, it is not surprising that the residual motion shows same magnitudes of a velocity comparable to that of the internal motion of the maser spots. 

We investigate a position of the central star and try to show it in figure~\ref{fig_maserdistribution}. 
We adopted three independent methods to estimate the stellar position. 
At first, we used data from our VLBI observations. 
Since the distribution of the maser spots was aligned along one standing out direction, north to south, we adopted a simple procedure in the first estimation. 
By averaging the distribution of all the maser spots in figure~\ref{fig_maserdistribution}, we obtained 
($-4.04\pm6.77$, $-14.54\pm30.58$) mas as a stellar position and presented it as a cross with solid line. 
Error of the position was derived from standard deviation of the maser spots. 
In the next estimation, we considered the internal motions of each maser spot. 
If all the maser spots move radially outwards from the central star, internal motion vectors of individual maser spots should intersect at a common origin. 
Based on this idea, we calculated all intersection points made by the internal motion vectors. 
Out of all 253 intersection points we excluded 28 points which were located beyond the distribution area of the maser spots. 
Then, using residual 225 intersection points, the average of these points was obtained to be ($-1.45\pm8.32$, $-23.56\pm20.45$) mas which is indicated as a cross with one-dotted chain line in figure~\ref{fig_maserdistribution}. 
Like the first estimation, a position error was derived from standard deviation of the intersection points from the average. 
In the last estimation, we used coordinate and proper motions in Gaia DR2 catalog. 
Using a coordinate of BX~Cam in Gaia DR2, we calculated position at the time of our VLBI observation in 10 February 2012. 
Then we obtained ($-0.27\pm3.78$, $-30.84\pm3.85$) mas as a stellar position and presented it as a cross with dotted line. 
Errors of the second position are estimated from errors of proper motions reported in the Gaia DR2 catalog. 
The three stellar positions estimated from independent methods fall on a central region of the whole maser distribution and show consistency within their error bars. 

Adopting a radial velocity of BX~Cam as 0 km\,s$^{-1}$, we can derive three dimensional expanding velocities of H$_2$O maser spots using $V_{\mathrm{LSR}}$ and internal proper motions. 
Obtained three dimensional velocities show a range of 12.67 km\,s$^{-1}$ to 18.68 km\,s$^{-1}$ for 23 maser spots. 
An average of the three dimensional velocities is 14.79 km\,s$^{-1}$ with a standard deviation of 1.40 km\,s$^{-1}$. 
This indicate that the H$_2$O masers coincide with an outflow with a quite uniform velocity. 
All maser spots detected in the VLBI observations of this work have blue-shifted LSR velocity with respect to the stellar velocity of 0 km\,s$^{-1}$. 
In a central region of figure~\ref{fig_maserdistribution}, we can find the bluest maser spots showing very small angular proper motions. 
For these bluest maser spots, velocities along the line of sight ($-$11 km\,s$^{-1}$ to $-$16 km\,s$^{-1}$) are significantly larger than the inplane velocities (1 km\,s$^{-1}$ to 4 km\,s$^{-1}$) perpendicular to the line of sight.  
Therefore, we can understand that the bluest maser spots clustered in the central region locate in front of the central star and moving toward us. 
From a comprehensive view of the spatial distribution and three-dimensional motions of the maser spots, we can see that there is a strong anisotropy of the circumstellar outflow in BX~Cam. 
Three collimated flows outwards from the central star are indicated from our VLBI observations. 

In the single-dish observation conducted in 19 November 2018, we detected a H$_2$O maser component at $V_{\mathrm{LSR}}$ of $+$8 km\,s$^{-1}$ which is redder than the systemic velocity of BX~Cam, however we did not detect them in our VLBI observations. 
Detection of the red-shifted maser components is expected to reveal the dynamical characteristics of circumstellar envelope of this source.  
For this purpose, additional VLBI observations are needed. 
More sensitive arrays like the combined array of the Korean VLBI Network (KVN) and VERA (KaVA)\footnote[1]{https://radio.kasi.re.kr/kava} 
or, 
the East Asian VLBI Network (EAVN)\footnote[2]{https://radio.kasi.re.kr/eavn} 
can contributes for better understanding of circumstellar matters that were not revealed in our VLBI observations. 

\section{Summary}
\label{sec_sum}
We measured an annual parallax of a Mira variable star BX~Cam through astrometric VLBI observations using the VERA. 
The parallax was obtained to be 1.73$\pm$0.03 mas which corresponds to the distance of 0.58$\pm$0.01 kpc. 
Parallactic oscillations in figure~\ref{fig_parallax} shows that angular motions of the seven maser spots were exactly traced by the successful astrometric observations with VERA. 
In the Gaia DR2 catalog, we can also find a parallax of 4.13$\pm$0.25 mas. 
They represent a large discrepancy of 240 \% compared to the parallax from VERA. 
Using both parallaxes, we estimated stellar luminosities as $L_{\ast}^{\mathrm{VERA}} = 4950\pm170 L_{\odot}$ and $L_{\ast}^{\mathrm{Gaia}} = 870\pm110 L_{\odot}$ from two parallaxes. 
General understanding about luminosities of Mira variable stars ($\sim10^3 L _{\odot}$ to $\sim10^4 L _{\odot}$) also support the validity of the VERA parallax. 
As a result, we rejected the parallax from Gaia DR2 and concluded that the parallax of 1.73$\pm$0.03 mas obtained from VERA is correct for BX~Cam. 
Assuming the parallax, we revealed kinematics of circumstellar masers in BX~Cam. 
We found 73 H$_2$O maser spots in 30 au $\times$ 80 au extent. 
We obtained the systemic proper motion of BX~Cam as 
(${\mu_{\alpha}}\cos{\delta}^{\mathrm{sys}}$, ${\mu}_{\delta}^{\mathrm{sys}}$)$_{\mathrm{VERA}}$ $=$ 
(13.48$\pm$0.14, $-$34.30$\pm$0.18) mas\,yr$^{-1}$. 
By averaging three dimensional velocity of the 23 H$_2$O maser spots, we obtained an expanding velocity of 14.79 km\,s$^{-1}$ with a standard deviation of 1.40 km\,s$^{-1}$. 
The stellar position was also estimated to be at the central region of the whole extent of the maser spots. 
Our VLBI observations suggest that the H$_2$O masers coincide with an outflow with a quite uniform velocity, however the flows are collimated and show strong anisotropy. 

\section*{Acknowledgement}
Data analysis were in part carried out on common use data analysis computer system at the Astronomy Data Center, ADC, of the National Astronomical Observatory of Japan.
This work has made use of data from the European Space Agency (ESA) mission {\it Gaia} (\url{https://www.cosmos.esa.int/gaia}), processed by the {\it Gaia} Data Processing and Analysis Consortium (DPAC, \url{https://www.cosmos.esa.int/web/gaia/dpac/consortium}). 
Funding for the DPAC has been provided by national institutions, in particular the institutions participating in the {\it Gaia} Multilateral Agreement.

{}
\end{document}